# DeepQuali: Initial results of a study on the use of large language models for assessing the quality of user stories

Adam Trendowicz[1], Daniel Seifert[1], Andreas Jedlitschka[1], Marcus Ciolkowski[2], and Anton Strahilov[3]

[1] Fraunhofer Institute for Experimental Software Engineering IESE,
Fraunhofer Platz 1, 67663 Kaiserslautern, Germany
```
{adam.trendowicz, daniel.seifert,
andreas.jedlitschka}@iese.fraunhofer.de
```
[2] QAware GmbH, Aschauer Str. 30, 81549 München, Germany
```
marcus.ciolkowski@qaware.de
```
[3] let's dev GmbH & Co. KG, Alter Schlachthof 33, 76131 Karlsruhe, Germany
```
anton.strahilov@letsdev.de
```

**Abstract.**
Generative artificial intelligence (GAI), specifically large language models (LLMs), are increasingly used in software engineering, mainly for coding tasks. However, requirements engineering - particularly requirements validation - has seen limited application of GAI. The current focus of using GAI for requirements is on eliciting, transforming, and classifying requirements, not on quality assessment. We propose and evaluate the LLM-based (GPT-4o) approach "DeepQuali", for assessing and improving requirements quality in agile software development. We applied it to projects in two small companies, where we compared LLM-based quality assessments with expert judgments. Experts also participated in walkthroughs of the solution, provided feedback, and rated their acceptance of the approach. Experts largely agreed with the LLM's quality assessments, especially regarding overall ratings and explanations. However, they did not always agree with the other experts on detailed ratings, suggesting that expertise and experience may influence judgments. Experts recognized the usefulness of the approach but criticized the lack of integration into their workflow. LLMs show potential in supporting software engineers with the quality assessment and improvement of requirements. The explicit use of quality models and explanatory feedback increases acceptance.

## 1    Introduction

Software engineering (SE) practitioners are increasingly recognizing the potential of Generative AI (GAI) to automate software development tasks [19], which have traditionally been labor-intensive, error-prone, and reliant on human expertise. Automation in SE was considered challenging due to the creative and unstructured nature of artifacts such as requirements specifications, design models, source code, test cases, and documentation. GAI's ability to process unstructured data and interpret context has created new opportunities, quickly adopted by SE research and practice [19].

As a result, the use of GAI – especially Large Language Models (LLMs) – is rapidly expanding across all phases of the software development lifecycle (SDLC). Most SE use cases for LLMs focus on coding, such as code generation and defect detection. In contrast, requirements engineering (RE) has received less attention despite its unstructured artifacts and lack of labeled data, which make it ideal for a GAI use case.

LLM use cases in RE [24] currently address constructive stages like elicitation, analysis, and specification [28]. Requirements validation, which assesses the quality of requirements specifications, remains the least automated RE stage. Automated validation is especially important in agile software development (ASD), where user stories change frequently and cycles are short.

Contribution: We evaluate DeepQuali, an LLM-based approach for assessing user story quality using explicit quality models. DeepQuali provides numeric quality scores, descriptions of quality issues, explanations, and improvement suggestions. The evaluation in two small software development companies focuses on user feedback regarding the quality assessments and the overall usefulness of DeepQuali in ASD projects.

The remainder of this paper is organized as follows: Section 2 reviews background and related work on requirements quality assurance (QA), including GAI applications. Section 3 introduces DeepQuali for assessing user story quality. Section 4 outlines the evaluation methodology. Section 5 summarizes and discusses the evaluation results. Section 6 addresses threats to validity. Section 7 concludes the paper.

## 2     Background and Related Work

**Quality Assessment of Software Requirements:** High-quality requirements are essential for successful software projects: They improve stakeholder communication, ensure that needs are met, support user-oriented testing, and enable reliable planning and effective scope control. Poor requirements can lead to costly rework, especially if detected late in the SDLC. Various constructive and analytical methods have been proposed to ensure quality requirements. Constructive methods "build quality in" during development include structured interviews, focus groups [4], prototyping and mockups [29], and formal specification [5]. Analytical methods "test quality in" by evaluating requirements through inspections and walkthroughs [21], automated checks against predefined rules [7], ambiguity detection6, traceability checks [27], and formal modeling [5]. Despite diverse tool-support [24], RE remains complex and time-consuming due to communication challenges and the unstructured nature of requirements.

**Using LLMs for Requirements Quality Assurance:** Machine learning (ML) and natural language processing (NLP) have been successfully used to automate requirements development [16]. Therefore, recent advances in LLMs have almost immediately led to their use in RE, mainly for the first three stages of the requirements process [28]: elicitation, analysis, and specification.

Elicitation involves understanding stakeholder needs, either directly through stakeholder or AI-agent dialogues [8], or indirectly by analysing user-generated content such as feedback, reviews, and social media [22]. Analysis and specification are the most automated RE phases [24]. LLMs support interpreting [23], classifying [14], prioritizing [22], and linking [11] requirements. Classification includes distinguishing functional/non-functional requirements [31] or identifying quality aspects like usability, performance, and security [1]. For specification, LLMs help transform natural language requirements into structured or formal formats [6]. The least automated RE phase is validation. It checks if requirements meet stakeholder needs [24], with most solutions based on NLP and deep learning [3]. These methods detect quality deficits [13, 9], however, typically very specific ones, such as ambiguity [3], indicated by syntactical or lexical elements of text. The Quality User Story framework [18] uses 13 criteria to assess user story quality in terms of syntax, pragmatics, and semantics. Recently, LLM-based approaches have been proposed for assessing and improving requirements quality. Typical scenarios involve using LLMs to identify quality issues via binary classification [9] or interval ratings [15], and to suggest improvements [18]. Some studies involved human experts evaluating the effectiveness and satisfaction with LLM-corrected user stories. Most evaluations compare LLM assessments to expert "ground truth," either by injecting synthetic deficits [22] or identifying real issues [15, 30, 17]. Evaluation metrics include agreement rates [30], interrater reliability [30], Cohen's coefficients [10], and ML metrics like precision, recall, and F-score [9, 17]. Some approaches use LLMs to generate feedback on detected issues, e.g., [15] found ChatGPT's feedback aligned well with human experts, outperforming CodeLlama in accuracy and consistency.

**Research Need:** Initial attempts have been made to use LLMs for requirements quality assessment. Further research is needed to improve these solutions and provide evidence of their practical usefulness and user acceptance. The current focus on predefined quality aspects (e.g., consistency, ambiguity, completeness, correctness) neglects practitioners' needs to define custom quality criteria, understand problems and causes, and get concrete solutions. Evaluations mainly rely on statistical comparisons with expert assessments, using standard ML accuracy measures. While statistical accuracy is important for acceptance, evaluations should also involve end users to get feedback, e.g., on perceived usefulness and acceptance.

## 3  LLM-Based Assessment of User Story Quality

To address research needs, we propose "DeepQuali", an LLM-based approach for assessing the quality of user stories. We use an existing LLM, i.e., GPT-4o with a custom-defined set of criteria in a structured and explainable manner. Fig. 1 illustrates the basic components of the approach. With DeepQuali, we envision three major benefits:

I) The explicit consideration of a *quality model*, represented, .e.g., by a quality standard (e.g., ISO/IEC 29148 [12]), industrial (e.g., INVEST [26], DoR [20]) or company-specific quality guidelines. II) A comprehensive *output* of the quality assessment, including quality ratings together with their explanation, descriptions of specific quality deficits associated with reference to the quality given guidelines, and corresponding improvement recommendations. III) Structured output assessments according to the predefined *output structure* template.

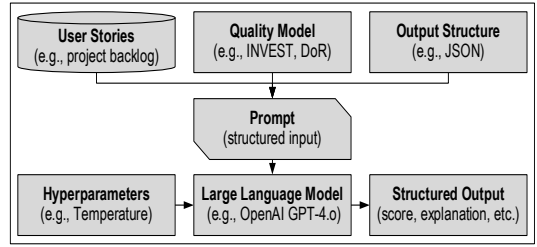

**Fig. 1.** DeepQuali approach

## 4     Evaluation Methodology

The study's **objective** was to evaluate the practical usefulness and user acceptance of LLM-based quality assessment of user stories in the context of industrial ASD projects from the viewpoint of intended users (i.e., agile software engineers). Specifically, this early evaluation aimed at exploring the practical applicability of DeepQuali, identify its deficits (critical feedback), explain root causes of the critique and identify improvement potentials. From this objective, we derived three **research questions**:

- **RQ1**: To what extent are the quality assessments estimated by DeepQuali consistent with those of human experts?
- **RQ2**: How do intended users evaluate the user story quality assessments provided by DeepQuali?
- **RQ3**: What is the level of acceptance of DeepQuali by intended users?

RQ1 refers to statistical accuracy of the quality assessments, RQ2 to usefulness of the quality assessments from the users' perspectives, and RQ3 to users' overall perceived acceptance of DeepQuali.

### 4.1     Study Context, Constraints, Data and Involved People

DeepQuali and its evaluation emerged from a collaborative research project involving two German SMEs specialized in software development. We investigated specific usage scenarios and stakeholders' needs regarding the requirements quality assessment in the **context** of ASD. We developed and evaluated the approach on sample data (user stories) and with the support of software engineering experts from two ASD projects, one from each of the involved companies. The projects were different regarding their application domain: one project developed a portal with online health courses (e.g., exercises) while the other developed a vehicle validity data management system.

A major **constraint** was data security and privacy. Sample user stories included reference information (meta data) that included sensitive data, which had to be removed before we could use it. Because of its supplementary character, its removal had no influence on the study results. Another constraint was the availability of subject-matter experts, who took the role of the intended users for DeepQuali. The study design had to be adapted to the time constraints of the experts involved. Therefore, we divided the evaluation into direct and indirect interaction with experts, i.e., on-site workshops and online surveys. A further constraint was the immaturity of the DeepQuali user interface, as it was in the initial development. Therefore, during the evaluation the users were guided through the application.

**Data** used in the study consisted of sample user stories provided by the involved companies. The study involved the following **people**:

- Experts: Four software engineers, two from each company, who were familiar with the project and user stories. They took the role of intended users of DeepQuali. They had different levels and areas of expertise (cf. Table 1).
- Researcher developed DeepQuali and designed and conducted the evaluation study.

**Table 1.** Characteristics of involved experts.

| | Company-1 | | Company-2 | |
|---|---|---|---|---|
| **Characteristic** | *Expert 1* | *Expert 2* | *Expert-1* | *Expert-2* |
| Role | SW engineer | SW Architect | Head of SE | Senior SE |
| Years in current organization | 2 | 4 | 4 | 11 |
| Years at current position | 2 | 13 | 4 | 7 |
| Years in Agile projects | 2 | 7 | 18 | 11 |
| Years specifying user stories | 2 | 7 | 16 | 7 |
| Familiar requirements quality assessment frameworks | None | None | QUS, INVEST | INVEST |
| Familiarity of AI principles | Moderate | Very | Very | Somewhat |
| Practical experience in using GAI | Extensive use | Frequent use | Frequent use | Frequent use |
| GAI methods used | ChatGPT, Github Copilot | ChatGPT, Github Copilot | Chat GPT, self-hosted LLMs | ChatGPT |

### 4.2 Study Procedure and Methods

Fig. 2 includes two development phases, which are related to the evaluation, yet not part of the evaluation itself: *Data acquisition and Development of DeepQuali.* In the first phase, we identified appropriate ASD projects and collected example user stories from the projects' backlogs. In the latter, we developed and optimized DeepQuali over several train-test iterations based on the deviations between obtained and expected (ground-truth) quality assessments. Additionally, we used the experience we gained while creating an LLM-based approach for inspecting software requirements [22]. The core evaluation procedure consisted of six steps:

**(1) Data selection** aimed at selecting and preparing a representative set of user stories for the evaluation. Because the evaluation involved experts from the participating companies, whose availability was limited, we reduced the sample size to five user stories. We asked experts to select the stories with perceived low, medium, and high level of complexity and quality. At this point, we left the definition of complexity and quality characteristics to the expert. We prepared the selected user stories for the evaluation, i.e., we anonymized and transformed them into a consistent JSON format.

**(2) Labelling survey** aimed at obtaining ground truth data. Experts rated the quality of selected user stories using a four-point interval scale on seven quality criteria: six criteria defined in INVEST and a summative Ready-to-Implement (RTI) criterion. The purpose of RTI was to obtain an "aggregated" concluding decision about the extent to which a given user story is ready to be released for implementation. To validate reliability of experts' ratings, each criterion was represented by three (INVEST) or four (RTI) statements shown in Table 2, each representing a sub-aspect of the corresponding high-level quality criterion as defined in the literature, e.g., [26].

Experts rated each statement with *Strongly disagree*, *Disagree*, *Agree*, or *Strongly agree*. To indicate equidistant intervals between the defined agreement ratings, we explicitly associated a 4-point interval scale (1 is the lowest and 4 the highest quality level) with them. We deliberately used a rating scale without a neutral point to force experts to take a clear position, thus avoiding the central tendency bias.

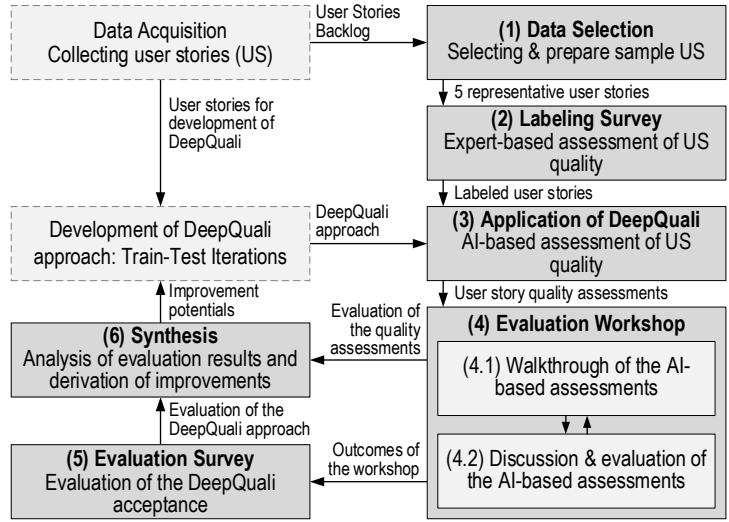

**Fig. 2.** Evaluation procedure.

**(3) Application of DeepQuali** aimed at assessing the quality of the selected user stories. We executed the LLM (GPT-4o) with parameters that have been optimized during the development of the approach. The parameters included:

- <u>LLM and its execution parameters,</u> including temperature, seed, maximal number of tokens, stop criteria as well as presence and frequency penalty (as listed in the outcomes of the model's application in ). During the initial model development, GPT-4o showed the best performance. In further iterations, we observed that the most consistent and reliable results were obtained at a temperature equal to zero (lowest allowed creativity of the model). At higher temperatures the model tends to provide inconsistent assessments for the same parameters and modifies input user stories; for instance, identified problems referred to a content that was not present in the original user story. Additional advantage of the OpenAI models is the support for structured outcomes "out-of-the-box". In the future, DeepQuali may adapt alternative LLMs.
- Prompts, including a system and a user prompt (cf. Table 3). The system prompt explains the LLM's basic assignment and its context. The user prompt specifies a specific task, e.g., user story's quality assessment on INVEST criteria or assessment of the readiness to implement.

Table 2. Labelling survey: quality criteria.

| | |
|---|---|
| **User story is independent.** | The user story describes a single, specific task or feature. |
| | The user story can be implemented independently of other user stories. |
| | The user story has minimal dependencies on other stories or external factors. |
| **User story is negotiable.** | The user story is open to discussion and refinement. |
| | The requirements and details of the user story are flexible and adaptable. |
| | The story can be modified or adjusted as more information becomes available. |
| **User story is valuable.** | The user story provides value to the end-user or customer. |
| | The user story is aligned with the project's goals and objectives. |
| | The user story delivers a tangible benefit or solves a specific problem. |
| **User story is estimable.** | The user story can be estimated with a reasonable degree of accuracy. |
| | The requirements and scope of the user story are clear and well-defined. |
| | The development team can provide a rough estimate of the time and effort required to implement the story. |
| **User story is small.** | The user story is small and manageable. |
| | The user story can be implemented within a single sprint or iteration. |
| | The user story has a limited scope and a clear, focused objective. |
| **User story is testable.** | The user story can be tested and verified. |
| | The user story acceptance criteria are clear and well-defined. |
| | The development team can write automated tests or create a test plan for the user story. |
| **User story is ready to implement.** | The user story fulfills acceptance criteria: Clear and understandable, consistent, validated, and realistic. |
| | The effort for the user story is estimated as story points. |
| | The user story description is sufficiently detailed. |
| | The user story description adheres to the predefined user story template. |

We forced the LLM to provide the output as a JSON by using the OpenAI API response format parameter. The output consists of several elements: the assessed user story, the LLM used and its execution parameters, and the quality assessments for each criterion defined. Each assessment encompasses: quality score (on the scale 1-worst to 4-best), explanation of the assessment, and specific problems identified (each problem description includes explanation, severity, and suggested solutions). The exact content of quality assessments depends on the specific quality criteria and tasks defined in the input prompt:

- INVEST criteria: The LLM has less freedom when assessing a user story in that it must consider INVEST criteria, although the criteria are not defined explicitly; it must "find out" the definition by itself. Problems: The LLM must focus on negative feedback for the predefined INVEST criteria.

- Ready to implement (RTI): The LLM was relatively free in the assessment of whether the user story is ready to be implemented. It had to consider the results from its assessments on the INVEST criteria.
- Custom Definition of Ready (DoR): The LLM must consider explicitly specifically defined DoR criteria. In our study these were provided by Company-1 only.

Table 3. Prompts used in the evaluation.

**System prompt**: You are an experienced software engineer with deep knowledge in agile practices and user story evaluation. You evaluate user stories on whether they are ready to be implemented. You respond in a given JSON schema.
**User prompt**:
- INVEST: I will provide you with a user story below. Your task is to carefully read through the user story and evaluate it using the INVEST criteria: {custom_prompt}. Here is the user story in JSON: {user_story_json}. Please stick to the given JSON schema.
- Ready to Implement: I will provide you with a JSON of a user story evaluation against the INVEST criteria. Your task is to carefully read through the evaluation and explain and classify whether the user story is ready to be implemented or not. First, provide an explanation and then give a binary classification whether the user story is ready to be implemented. Here is the user story evaluation: {invest_json}.
- Custom Definition of Ready: I will provide you with a user story and a custom definition of ready (DoR) below. Your task is to carefully read through the user story and evaluate it using the custom DoR. Here is the custom DoR: Definition of Ready: {dor} Here is the user story in JSON: User story {user_story_json}.

**(4)** During **Evaluation workshops**, we showed the intended users how to work with DeepQuali. The researcher walked the experts through DeepQuali for the selected sample user stories one by one. For each user story:

*(4.1) Group discussion and feedback*: Experts asked clarification questions regarding the application process and outcomes. Then the researcher moderated a group discussion on DeepQuali's strengths and weaknesses. The researcher collected and documented the experts' feedback. Discussed aspects include: (1) clarity and completeness of the explanations and problems, (2) compliance of the explanations and problems with the predefined criteria definitions (INVEST and DoR), (3) consistency of the explanations with one another as well as with the corresponding numeric quality and severity scores, and (4) consistency of the assessments with the experts' assessments and expectations.

*(4.2) Individual feedback survey*: Experts filled out a survey, in which they evaluated the quality assessments provided by DeepQuali on several criteria (Table 5). The assessments included quality scores and their explanations on INVEST and RTI criteria as well as quality Problems and their criticality. In Company-1, experts evaluated additionally the DeepQuali assessments on company's custom-defined Definition-of-Ready (DoR) criteria (Table 4).

Table 4. Definition of ready (DOR) criteria

| Criterium | Definition |
|---|---|
| Acceptance Criteria (ACs) | Clearly defined, consistent, validated by stakeholders, and realistic. |
| Effort Estimation | Includes a size metric (e.g., XL-L-M-S-XS) or Story Points, mapped to hours agreed upon by the team. |
| Description | Sufficiently detailed for implementation, with proportional effort for specification and implementation. Minor kick-off discussions are acceptable. |
| Template Completion | Must follow the story template, with irrelevant elements optionally skipped. |
| Story Template | "As a [ROLE], I want [GOAL] so that [BENEFIT]" |
| Design Background | Provides context such as interfaces, data models, or frameworks. |
| Prerequisites | Lists external interfaces, infrastructure, and internal preparations. |
| Steps to Achieve Goal | Non-binding tasks outlined as bullet points, optionally with useful hints. |
| Out of Scope | Clarifies excluded tasks or linked follow-up stories. |
| Criteria for Goal Achievement | Includes data-based test cases, test data, and validation patterns. |
| Test Scenario | Defines click-routes and explicit data for verification in INT and PROD. |
| Hints | Optional additional information to aid implementation. |
| Contacts | Specifies responsible persons or knowledge holders. |

**Table 5.** DeepQuali outcomes' evaluation criteria

| Criterium | Definition (DeepQuali …) |
|---|---|
| Accurate | … correctly identifies actual quality deficits, meaning, it does not identify issues where there are none (false positives) and it does not miss actual quality issues (false negatives). |
| Complete | … identifies all significant quality deficits. |
| Relevant | … provides output that is directly related to the user story and the project context in question. |
| Explainable & Clear | … explains in a clear, concise, and specific manner why it flagged a specific part of a user story as deficient. |
| Actionable | … provides clear guidance on how to improve the user story, specifically it prioritizes the issues based on their impact on the overall quality of the user story and on their impact on the quality of software. The suggestions should be practical and actionable. |
| Consistent | … provides consistent feedback for similar user stories or quality deficits. |
| Context-driven | … demonstrates an understanding of the project's context, domain, and any specific requirements or constraints when evaluating user stories. |
| Quality-conform | … assessments align with established quality criteria for user stories and consistently apply the same standards and criteria across different user stories. |

**(5) Evaluation survey** aimed at evaluating the intended users' acceptance of DeepQuali. For this purpose, we designed a survey based on the Unified Theory of Acceptance and Use of Technology (UTAUT-2) [25]. However, because the DeepQuali approach was new and still immature (e.g., a fully functional software tool cannot be used by experts), we have removed the following UTAUT-2 constructs:

- *Social influence*, because due to the setting, there was no social influence possible,
- *Hedonic motivation*, because users cannot use the tool by themselves,
- *Price value*, because the tool is intended to be free of charge,
- *Experience and habit*, because users have not seen or used the approach before.

and adapted the remaining constructs to DeepQuali as technology.

**(6) Synthesis** aimed at analyzing evaluation outcomes and deriving improvement potential. Specifically, we performed the following data processing and analyses:

- We transformed agreement-scale data to a numerical, interval scale (1-Strongly disagree to 4-Strongly agree).
- We aggregated the survey evaluations on individual sub-aspects of INVEST and RTI criteria (steps 2 and 4) using the median.
- We analyzed consistency of the user story quality assessments between experts (step 2). For this purpose, we used Spearman's rank correlation (Rho) and Kendal's coefficient of concordance (Tau).
- RQ1: We analyzed quantitatively the consistency (agreement) between DeepQuali's and the experts' quality assessments. We computed the absolute deviation between numerical ratings. For comparability with related work, we also evaluated the consistency of ratings using the metrics known from the analysis of the predictive performance of machine learning classification models. Therefore, we treated quality rating as nominal values (classes) and calculated the following metrics: accuracy, precision, recall and F1-score. We treated experts' ratings as ground truth and those by DeepQuali as predictions.
- RQ2: We quantitatively analyzed experts' evaluations of individual assessments provided by DeepQuali.
- RQ2: We qualitatively analyzed the unstructured feedback provided by the experts during walkthrough workshops. For this purpose, we computed basic descriptive statistics and visualized the results using box plots.
- RQ3: We quantitatively analyzed the experts' acceptance ratings. For this purpose, we computed basic descriptive statistics and visualized the results using box plots.
- We derived potential improvements and future research regarding the application of GAI/LLMs for assessing and improving the quality of software requirements.

Because of the small sample size (three to five user stories and four experts), we did not calculate tests of statistical significance.

## 5      Results and Discussion

We report the results of the evaluation study and discuss them in the light of our research questions. Due to time constraints, only three out of the five sample user stories were considered during the workshop with experts at Company-1. This may influence the perceived usefulness of user stories' quality assessment provided by DeepQuali (RQ2) and the acceptance of the approach (RQ3) by experts from Company-1, compared to experts from Company-2 where the evaluation was performed on all five user stories.

**RQ1: To what extent quality assessments estimated by DeepQuali are consistent with human experts' assessments?**

*1) Agreement between experts.*

Table 6 presents the coefficients of agreement between two experts from each company, regarding their quality assessments of the sample of five user stories. We calculated coefficients for each quality criterion based on the ratings of its sub-aspects' consistency. Additionally, we calculated an overall an agreement for *All* criteria on all sub-aspects. Missing values are due to the inability to compute the coefficients for constant values (the same ratings) provided by at least one of the experts, on all quality aspects considered.

Results indicate rather weak agreement between experts. They agree with each other to the greatest extent when assessing quality of user stories on the INVEST criterion *Small* and when assessing the overall readiness of a user story to be implemented (RTI criterion). Experts disagree to the greatest extend – however only in Company-1 – when assessing quality of user stories on the INVEST's criteria *Negotiability* and *Testability*.

**Table 6.** Agreement of users' assessments. Values in bold font are significant at p = 0.05, due to small sample size they have not any statistical power

|  |  | Independent | Nogotiable | Valuable | Estimable | Small | Testable | RTI | All |
|---|---|---|---|---|---|---|---|---|---|
| Company-1 | *Tau* | 0,47 | 0,19 | 0,44 | 0,41 | **0,49** | 0,31 | 0,39 | **0,36** |
| Company-1 | *Rho* | 0,50 | 0,20 | 0,49 | 0,41 | **0,55** | 0,35 | 0,42 | **0,40** |
| Company-2 | *Tau* | 0,31 | - | - | 0,34 | **0,54** | - | 0,54 | **0,46** |
| Company-2 | *Rho* | 0,32 | - | - | 0,35 | **0,55** | - | 0,61 | **0,50** |

One potential cause of the observed inconsistency of experts' assessments could be their areas and levels of expertise. In both companies, the two experts differed widely with respect to their area of expertise (role in ASD projects) and seniority.

*2) Consistency between Experts and DeepQuali*

Fig. 3 illustrate deviations between DeepQuali's and experts' assessments of user stories' quality. The deviation is the value of experts' assessment minus the value from the DeepQuali assessment. Consequently, a negative deviation means that DeepQuali assessed the quality of a given user story higher than expert, and vice versa. The DeepQuali-Experts assessments' consistency differed between companies. For Company-1 the lowest overall deviation was on the *Estimable* criterion although it was not consistent between the two experts; the highest deviations were on the *Negotiable* and *Testable* criteria, whereas experts assessed quality of sample user stories lower than DeepQuali. Furthermore, the expert with larger experience (E1) tended to be more critical (provide lower quality assessments) than the second expert (E2) and then DeepQuali. For Company-2, the lowest overall deviation was observed on the *Valuable* criterion; the highest deviations were observed on the *Estimable* and *Testable* criteria, where experts' assessments on these aspects were lower than DeepQuali's. The only criterion on which all expert assessments were higher than DeepQuali's was *Small*. One possible reason might be that DeepQuali expected significantly smaller user stories than it used to be common in the two companies. Another possible explanation might be that DeepQuali was misled by detailed information included in a user story or structure of a user story. For example, user stories of Company-2 included an explicit list of acceptance criteria; if a user story defined a multiple acceptance criteria, DeepQuali suggested that "although being relatively small" the user story should be split into smaller tasks. Similarly, user stories of Company-1 explicitly defined lists of detailed steps or tasks associated to a functionality – this might have led DeepQuali to propose a split of the user story. Therefore, future improvements should incorporate any company-specific standards for formulating and structuring user stories.

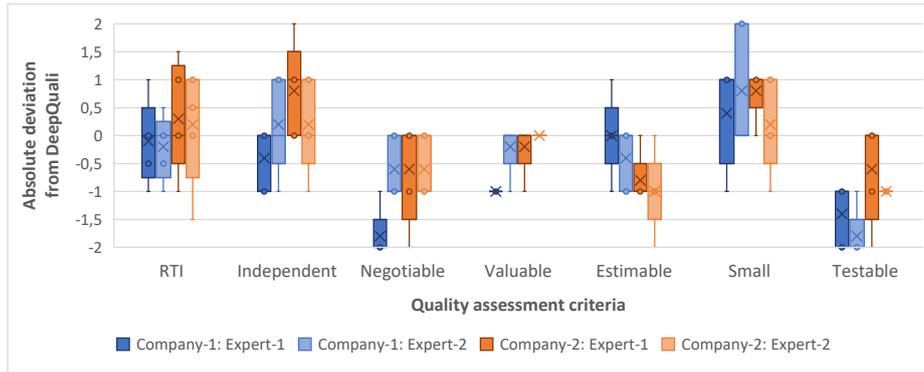

**Fig. 3.** Deviations between expert and DeepQuali quality assessments

Analysis of the classification accuracy, precision, recall and F1-score (Table 7) confirmed the rating deviations illustrated in Fig. 3. Wherever the deviation between quality ratings provided by an expert and by DeepQuali are large, the classification performance is low. However, such simple relations do generally not exist because absolute deviations represent the size of interval between ratings whereas classification performance ratings represent the matches of nominal values (in a confusion matrix).

**Table 7.** Classification performance metrics

|  | Criteria | Expert-1 | | | | Expert-2 | | | |
| --- | --- | --- | --- | --- | --- | --- | --- | --- | --- |
|  |  | *Accuracy* | *Precision* | *Recall* | *F1-score* | *Accuracy* | *Precision* | *Recall* | *F1-score* |
| Company-1 | Independent | 0,60 | 0,36 | 0,60 | 0,45 | 0,40 | 0,16 | 0,40 | 0,23 |
|  | Negotiable | 0,00 | 0,00 | 0,00 | 0,00 | 0,40 | 0,85 | 0,40 | 0,40 |
|  | Valuable | 0,00 | 0,00 | 0,00 | 0,00 | 0,80 | 0,85 | 0,80 | 0,78 |
|  | Estimable | 0,60 | 0,60 | 0,60 | 0,60 | 0,60 | 0,80 | 0,60 | 0,57 |
|  | Small | 0,20 | 0,20 | 0,20 | 0,20 | 0,60 | 0,87 | 0,60 | 0,60 |
|  | Testable | 0,00 | 0,00 | 0,00 | 0,00 | 0,00 | 0,00 | 0,00 | 0,00 |
|  | RTI | 0,60 | 1,00 | 0,60 | 0,75 | 0,60 | 0,60 | 0,60 | 0,60 |
| Company-2 | Independent | 0,40 | 0,40 | 0,40 | 0,40 | 0,40 | 0,53 | 0,40 | 0,46 |
|  | Negotiable | 0,60 | 0,67 | 0,60 | 0,58 | 0,40 | 1,00 | 0,40 | 0,57 |
|  | Valuable | 0,80 | 0,64 | 0,80 | 0,71 | 1,00 | 1,00 | 1,00 | 1,00 |
|  | Estimable | 0,20 | 0,40 | 0,20 | 0,27 | 0,20 | 0,30 | 0,20 | 0,24 |
|  | Small | 0,20 | 0,40 | 0,20 | 0,27 | 0,40 | 0,87 | 0,40 | 0,42 |
|  | Testable | 0,60 | 0,36 | 0,60 | 0,45 | 0,00 | 0,00 | 0,00 | 0,00 |
|  | RTI | 0,40 | 0,87 | 0,40 | 0,42 | 0,40 | 0,73 | 0,40 | 0,46 |

**RQ2: How intended users evaluate the quality of user story quality assessments provided by DeepQuali?**

Table 8 shows experts' evaluations of the user story quality assessments by DeepQuali. Each cell contains an aggregated evaluation on a given aspect for all assessed user stories. Grey cells indicate the lowest median evaluations. Company-1 experts were most critical regarding the actionability, completeness and consistency of DeepQuali assessments on the INVEST and RTI criteria. In this case, the low actionability of the assessments was partially associated with their perceived incompleteness and inconsistency. Experts positively evaluated the way DeepQuali reported problems for specific INVEST criteria. Experts deviated consistently with respect to their evaluations of DeepQuali assessments on the custom-defined DoR quality criterion. The more experienced Expert-2 was consistently more critical in the evaluations. In their feedback, experts positively evaluated the explanations DeepQuali provided in addition to numerical ratings; experts considered the explanation as very helpful to comprehend assessments. Regarding the RTI criterion, experts wished to receive binary assessment, instead of four-point one, because it would help deciding quicker whether a user story is ready for implementation or not.

Company-2 experts evaluated outputs of DeepQuali higher than Company-1 experts. Like in Company-1, the more experienced Expert-2 (E2) was more critical than the less experienced Expert-1 (E1). Experts noted that assessments of individual user stories on the *Independent* criterium may not be meaningful without considering the larger context, e.g., relations to remaining user stories in the project backlog. Experts liked the explanations associated with quality ratings as they increased their understanding of DeepQuali's "reasoning". They also appreciated the assessments on RTI as a summary for easier decision making. However, they were confused by the user stories' quality assessments, where DeepQuali did not give the highest RTI, although it has not identified any problems on INVEST criteria. Experts got the impression that DeepQuali "searches" for problems even if there are none and in consequence provide biased quality ratings. Finally, experts criticized ambiguous formulations, such as "it may" or "it could", used by DeepQuali in explanations and problem descriptions; they expected more definitive assessments. Regarding problems, experts suggested assessing explicitly positive and negative aspects of user stories, instead of only problems.

**Table 8.** Median evaluations of the DeepQuali assessments' quality

|  | Company-1 (three user stories) | | | | | | | | Company-2 (five user stories) | | | | | |
|---|---|---|---|---|---|---|---|---|---|---|---|---|---|---|
|  | INVEST | | RTI | | Problems | | DOR | | INVEST | | RTI | | Problems | |
|  | E1 | E2 | E1 | E2 | E1 | E2 | E1 | E2 | E1 | E2 | E1 | E2 | E1 | E2 |
| Accurate | 3 | 2 | 3 | 2 | 3 | 3 | 3 | 2 | 4 | 3 | 4 | 4 | 3 | 3 |
| Actionable | 2 | 2 | 3 | 2 | 3 | 3 | 3 | 2 | 4 | 3 | 4 | 4 | 4 | 2 |
| Complete | 2 | 2 | 2 | 2 | 3 | 2 | 3 | 2 | 4 | 3 | 4 | 4 | 3 | 2 |
| Consistent | 2 | 2 | 3 | 2 | 3 | 3 | 3 | 2 | 3 | 3 | 3 | 3 | 3 | 3 |
| Context-driven | 3 | 3 | 3 | 3 | 3 | 2 | 3 | 1 | 4 | 4 | 4 | 4 | 4 | 4 |
| Explainable and Clear | 2 | 3 | 3 | 3 | 4 | 3 | 3 | 2 | 4 | 3 | 4 | 3 | 4 | 3 |
| Quality comform | 2 | 3 | 3 | 2 | 3 | 3 | 3 | 2 | 4 | 3 | 4 | 4 | 4 | 3 |
| Relevant | 3 | 2 | 3 | 3 | 3 | 4 | 3 | 2 | 4 | 4 | 4 | 4 | 4 | 4 |

**RQ3: What is the acceptance of DeepQuali by intended users?**

Users agreed on the potential usefulness of DeepQuali in ASD practice. Fig. 4 shows the results of the perceived acceptance survey. Experts agreed that DeepQuali is easy to use and has potential to improve their work performance and they expressed their willingness to use it. However, they criticized insufficient facilitating conditions for using DeepQuali in their daily work, e.g., an integrated tool seems to be a critical condition for the acceptance.

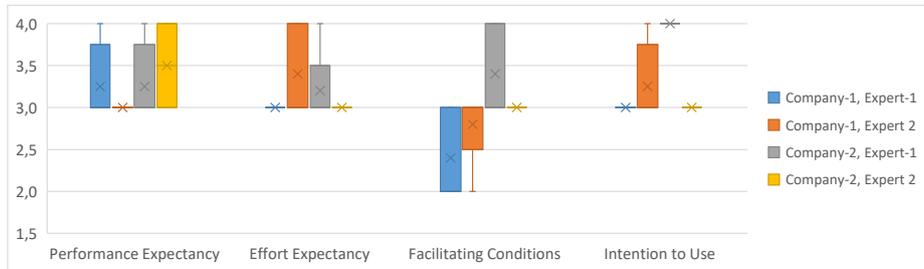

**Fig. 4.** Results of the survey on perceived user acceptance of DeepQuali.

## 6      Threats to Validity

*Construct validity* threats mainly stem from the immaturity of DeepQuali and the early stage of the study. Since potential users were guided through DeepQuali rather than using it independently, this affects the validity of the "user acceptance" construct.

*Internal validity* threats arise from limited control over study context characteristics that could influence results. E.g., experts' background and experience likely affected their opinions on DeepQuali's usefulness. We observed rating differences among experts with varying experience levels. Future studies should consider such confounders.

*Conclusion validity* threats are due to the study's limited scope. The small sample size—both in terms of experts and user stories—prevented isolation of influencing factors like differing interpretations of INVEST criteria or flaws in their definitions. Additionally, the probabilistic nature of the LLM affects replicability, although we minimized this risk by adjusting its parameters.

*External validity* threats relate to the study's restriction to two companies and two ASD projects. Broader studies in other contexts are needed to generalize the findings.

## 7      Summary and Conclusions

With DeepQuali, an LLM-based approach for assessing the quality of user stories in ASD, we address a gap in using LLMs for requirements QA. We present the results of a study evaluating DeepQuali, which took place in two small software companies, each providing a set of user stories and two subject matter experts. DeepQuali assessed user story quality using a defined quality model, specifically the INVEST criteria and custom Definition-of-Ready criteria. The tool provided quantitative ratings, explanations, descriptions of issues with specific quality criteria, and an overall readiness assessment. The evaluation included an analysis of DeepQuali's consistency with expert judgments, as well as experts' views on its practical usefulness and acceptance. While the results are not statistically significant, they are promising. Participants found DeepQuali supportive and provided feedback for further development. Experts valued the explanations, explicit consideration of quality criteria, and the summary rating for decision-making. However, they noted that DeepQuali sometimes identified problems that do not exist and suggested it should also highlight positive aspects and consider project context. They also agreed on DeepQuali's potential to improve their work and expressed willingness to use it, provided it is integrated into their specific work environments. For practitioners and researchers, our study offers initial evidence that LLMs can be applied to requirements QA. Practitioners can build on our results, while researchers might explore the use of alternative LLMs.

Future work will focus on improving DeepQuali based on feedback, extending its capabilities to assess multiple user stories and additional quality aspects such as completeness, consistency, redundancy, contradictions, and assessing coherence with epics. Finally, we will address the uncertainties of GAI and their management.

## Acknowledgments

Parts of this work were funded by the Federal Ministry of Research, Technology and Space (BMFTR) under grant no. 01IS23016D. We are very grateful to our participants.